\documentclass[aps, pra, twocolumn, notitlepage,superscriptaddress,nofootinbib]{revtex4-1}

\usepackage{natbib}
\usepackage{graphicx}
\usepackage{dcolumn} 
\usepackage{bm}
\usepackage{hyperref}
\usepackage{mathtools}
\usepackage{amssymb}
\usepackage{enumitem}
\usepackage{xcolor}
\usepackage{ulem}
\usepackage[mathlines]{lineno}
\usepackage[titletoc, title]{appendix}

\begin{document}

\title{Interaction-resistant metals in multicomponent Fermi systems}

\author{Andrea Richaud}
\email{arichaud@sissa.it}
\affiliation{Scuola Internazionale Superiore di Studi Avanzati (SISSA), Via Bonomea 265, I-34136, Trieste, Italy}
\author{Matteo Ferraretto}
\affiliation{Scuola Internazionale Superiore di Studi Avanzati (SISSA), Via Bonomea 265, I-34136, Trieste, Italy}
\author{Massimo Capone}
\affiliation{Scuola Internazionale Superiore di Studi Avanzati (SISSA), Via Bonomea 265, I-34136, Trieste, Italy}
\affiliation{CNR-IOM Democritos, Via Bonomea 265, I-34136 Trieste, Italy}

\date{\today}

\begin{abstract}

We analyze two different fermionic systems that defy Mott localization showing a metallic ground state at integer filling and very large Coulomb repulsion. 
The first is a multiorbital Hubbard model with a Hund's coupling, where this physics has been widely studied and the new metallic state is called a Hund's metal, and the second is a SU(3) Hubbard model with a patterned single-particle potential designed to display a similar interaction-resistant metal in a set-up which can be implemented with SU($N$) ultracold atoms. 
With simple analytical arguments and  exact numerical diagonalization of the Hamiltonians for a minimal three-site system, we demonstrate that the interaction-resistant metal emerges in both cases as a compromise between two different insulating solutions which are stabilized by different terms of the models. This provides a strong evidence that the Hund's metal is a specific realization of a more general phenomenon which can be realized in various strongly correlated systems.



\end{abstract}

\maketitle

\section{Introduction}
\label{sec:Introduction}

The Mott metal-insulator transition is one of the most striking manifestations of electronic correlations~\cite{Imada_RMP}. In a single-orbital Hubbard model at half filling (one particle per lattice site), carriers localize when the ratio between the Hubbard on-site repulsion $U$ and the hopping  $t$ exceeds a certain critical value $U_c/t$. 
In multiorbital models, this simple scenario is modified as $U$ is supplemented by interorbital terms, such as  the Hund's exchange coupling, $J$, which favours high-spin configurations~\cite{Haule_Kotliar,De_Medici_Janus,De_Medici_multiorbital_correlations}. 
For instance, an effective decoupling between the orbitals~\cite{De_Medici_out_lifting,De_Medici_multiorbital_correlations,De_Medici_Selective_Mott,
Yin2011,Fanfarillo_Electronic_correlations,Capone_Citro} opens the doors to orbital-dependent correlations and even to orbital-selective Mott phases where only some orbitals become Mott-localized~\cite{De_Medici_out_lifting,Vojta_Orbital_Selective,De_medici_orbital_selective,Ferrero_orbital_selective}, while the spin degrees of freedom are partially frozen~\cite{Spin-freezing}, anomalous responses are observed~\cite{Stadler_spin_orbital_separation}, and a peculiar spectral weight redistribution favours orbital-selective superconductivity~\cite{Fanfarillo2020}.

Turning back to the Mott transition, it has been shown that in a $N$-orbital Hubbard model at integer filling different from $N$, increasing $J/U$ (see below for the definition of the Hamiltonian) pushes $U_c/t$ to very high values~\cite{De_Medici_Janus} leading to a wide region of an interaction-resistant metal that has been labelled as a ``Hund's metal".
For the specific value $J/U=1/3$, the system remains metallic even in the limit of infinite $U$ and $J$. This scenario is particularly surprising, at least at first sight, because both $U$ and $J$ constrain the motion and are expected to localize the carriers. 
In Ref.~\cite{Isidori_Charge_disproportionation}, the resilience of the metallic solution for large interactions has been described in terms of the competition between two strongly correlated solutions, a high-spin Mott state favoured by $U$ and a disproportionated Hund's insulator favoured by $J$. 


In the last years, a new platform for the quantum  simulation of multicomponent Hubbard-like models has been realized by means of ultracold  $^{87}\mathrm{Sr}$~\cite{Strontium87}, $^{173}\mathrm{Yb}$~\cite{Ytterbium173}, and $^{6}\mathrm{Li}$ atoms~\cite{Ottenstein_Li_6,Huckans_Li_6}. In these atoms, the nuclear spin is essentially decoupled from the electronic degrees of freedom, thus providing one with $10$, $6$, or $3$ different spin flavors, and allowing for the quantum simulation of SU($N$)-symmetric Hubbard models, if loaded in deep optical lattices ~\cite{Gorshkov,Cazalilla,Pagano_1,Zhang,Pagano_2,Hofer,Cappellini,Scazza,Del_Re_Capone}.

Rather than a direct quantum simulation of the multiorbital Hubbard model with Hund's coupling, which is complicated by the need to define an orbital and a spin degree of freedom, we propose a different model which is expected to display a physics similar to the one of the Hund's metal, namely an SU($3$) system with a site-dependent single-particle potential characterized by a three-site pattern where two sites out of three have a lower energy with respect to the third. In this model the standard Mott insulator competes with a density wave pattern, similarly to the ionic Hubbard model on bipartite lattice.



In this work, we solve the two models by exact diagonalization in the smallest system that can capture the main physics of the two models, namely a three-site cluster, or ``trimer". The choice of this minimal lattice allows one to perform a full diagonalization of the Hamiltonian and to present thorough results for both the ground state and the thermal properties in an unbiased way. In particular, as we discuss in the following, the trimer is the smallest cluster which can host all the insulating solutions expected in the model.

For the multiorbital Hubbard model, we show that the trimer reproduces the most important features which have been uncovered in the thermodynamic limit using Dynamical Mean-Field Theory (DMFT)~\cite{Georges_Z_propto_D}, slave-spin mean-field theory~\cite{Capone_Citro}, and rotation-invariant slave bosons (RISB)~\cite{risb}. 
This agreement represents a cross validation of different approaches with different limitations. In fact all the above methods are defined in the thermodynamic limit, but they treat dynamical correlations only at a local level, while the exact diagonalization of the trimer obviously includes also non-local (nearest-neighbor) correlations, but it suffers from finite-size effects. 
The present results complement the previous literature with an analysis of the low-energy excitations and their relation with finite-temperature properties and with the investigation of nearest-neighbor correlations and their comparison with the on-site counterparts.

We highlight the similarity between the three-band Hubbard model and the three-flavor Hubbard model, which both display an interaction-resilient metal as a result of the competition between two different localization mechanisms. In both cases the metallic character is shown to be supported by the coexistence, within the same state, of atomic multiplets characteristic of the two competing localized solutions. 

The outline of our work is the following. In Sec. \ref{sec:Kanamori_trimer} we investigate the three-site three-orbital Hubbard model with Kanamori interactions. In Sec. \ref{sec:Hubbard_trimer} we analyze the different physical regimes offered by a three-flavor ultracold fermionic gas with a patterned potential. Sec. \ref{sec:Concluding_Remarks} is devoted to some concluding remarks.

\section{The Hubbard-Kanamori trimer}
\label{sec:Kanamori_trimer}
In this section we introduce the three-orbital Hubbard-Kanamori model on a trimer
\begin{equation}
\label{eq:H_Kanamori}
     H=-\sum_{ij,ab,\sigma} t_{ij}^{ab} d^\dagger_{ia\sigma}d_{jb\sigma} +\sum_j H_{\mathrm{int},j}
\end{equation}
where operator $d_{ia\sigma}^\dagger$ creates a spin-1/2 fermion with spin $\sigma=\uparrow,\downarrow$ in orbital $a=1,2,3,$ of site $i$. We assume three degenerate levels with orbital-diagonal hopping $t_{ij}^{ab}=t_{j,j+1}^a\delta_{ab}\delta_{ij}=:t\, \forall j,a$, i.e., we do not include hybridization between orbitals or crystal-field splitting. We expect our results to be robust with respect to terms breaking the symmetry between the orbitals, including different hopping, hybridization and crystal-field splitting as long as these terms are not comparable with the interaction terms. The robustness of the results is expected also in light of previous calculations for realistic electronic structures for iron-based superconductors \cite{YuSi,Lanata,De_Medici_Selective_Mott} which confirm the picture obtained for the symmetric model.

The local interaction  reads 
\begin{align}
\label{eq:H_int_Fock}
H_{\mathrm{int},j} & = U\sum_a n_{ja\uparrow}n_{ja\downarrow} 
+ (U-3J)\sum_{a<b, \, \sigma} n_{ja\sigma}n_{jb\sigma} \nonumber\\
& + (U-2J)\sum_{a\neq b} n_{ja\uparrow}n_{jb\downarrow}  
+ J\sum_{a\neq b} d^\dagger_{ja\uparrow}d^\dagger_{ja\downarrow}\,d_{jb\downarrow}d_{jb\uparrow} \nonumber\\
& - J \sum_{a\neq b} d^\dagger_{ja\uparrow}d_{ja\downarrow}\,d^\dagger_{jb\downarrow}d_{jb\uparrow},
\end{align}
where $U>0$ and $J>0$ are the standard Hubbard repulsion, and the Hund's exchange coupling, respectively. 

The interaction term (\ref{eq:H_int_Fock}) can be recast in the form 
\begin{equation}
\label{eq:H_int_n_L_S}
    H_{\mathrm{int},j} = \frac{U-3J}{2}\hat{n}_j(\hat{n}_j-1) -J\left(2\mathbf{S}_j^2 + \frac{1}{2}\mathbf{L}_j^2 - \frac{5}{2} \hat{n}_j\right)
\end{equation}
%
%
%
%
where $\hat{n}_j = \sum_{a\sigma}d^\dagger_{ja\sigma} d_{ja\sigma}$ counts the total number of fermions at site $j$, ${\mathbf{S}_j} = \frac{1}{2} \sum_{a,\,\sigma\sigma'} d^\dagger_{ja\sigma} \hat{\bm{\sigma}}_{\sigma\sigma'} d_{ja\sigma'}$ and ${\mathbf{L}_j} = \sum_{ab,\,\sigma} d^\dagger_{ja\sigma} \hat{\bm{\ell}}_{ab} d_{jb\sigma}$ represent the local spin and orbital angular momentum operators, being $\hat{\bm{\sigma}}_{\sigma\sigma'}$ the Pauli matrices and $\hat{\bm{\ell}}^{(a)}_{bc} = - i \epsilon_{abc}$ the generators of group $O(3)$. The form (\ref{eq:H_int_n_L_S}) is rather instructive, as it highlights the presence of two contributions: the on-site repulsion ($\propto U-3J$) and the exchange mechanism ($\propto J$) which favours primarily high-spin states and, as a second condition, high-orbital-angular-momentum states, thus realizing the first two Hund's rules. 

We consider a system with $L=3$ lattice sites (so that $i,j=1,2,3$) with periodic boundary conditions and a total of 6 particles, i.e. 2 particles on average per site. 
This very small system has proven to be an effective minimal lattice in the investigation of a number of physical phenomena, for both fermionic~\cite{Shiba_Pincus_Specific_heat,Ullrich_Trimer,Schilling_trimer,Aligia_trimer} and bosonic~\cite{Richaud_PRA_supermixing,Richaud_Sci_Rep_1,Sci_Rep_Zenesini,Richaud_NJP,Las_Phys_2018} systems.
To our purposes, the trimer system is particularly useful because it can host all the insulating solutions expected in our model (and in the second model we discuss in Sec. \ref{sec:Hubbard_trimer}). Furthermore, small clusters can be used as building blocks for quantum cluster theories such as cluster perturbation theory \cite{Cluster-Perturbation-Theory}, variational cluster approximation \cite{Potthoff} or cluster extensions of DMFT \cite{CDMFT,DCA}.

\subsection{Atomic multiplets}
\label{sub:Rise_and_fall}
We begin reviewing known results about the atomic limit ($t=0$) of the Hamiltonian (\ref{eq:H_Kanamori}) which are useful to understand the results and to compare them with the three-component model we propose in Sec. \ref{sec:Hubbard_trimer}.

\begin{figure}[h!]
    \centering
    \includegraphics[width=1\linewidth]{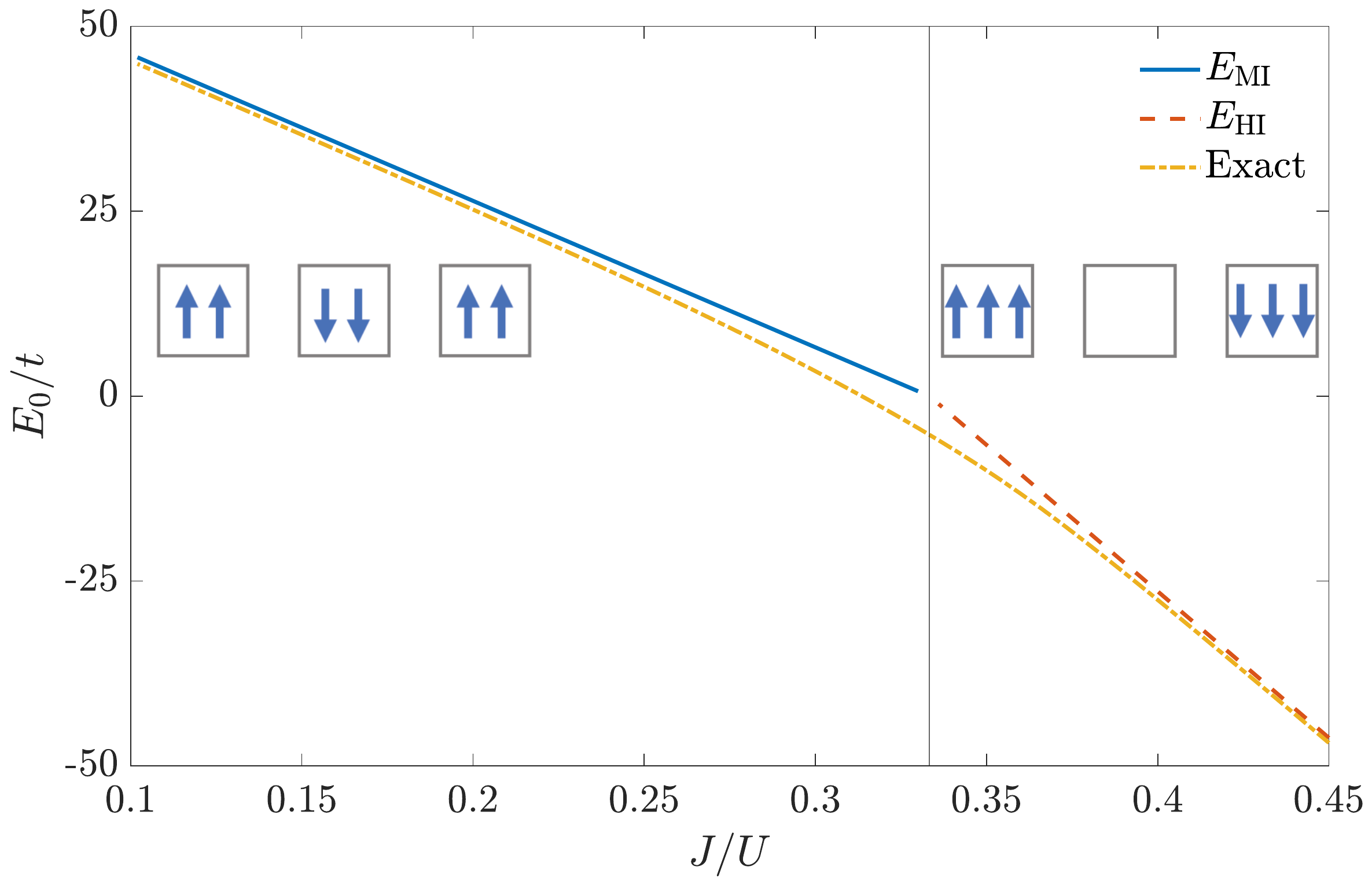}
    \caption{Ground state energy of a three-band Kanamori trimer hosting 6 particles as a function of $J/U$. Comparison between exact numerical diagonalization of Hamiltonian  (\ref{eq:H_Kanamori}) for $t=1$ and $U=22$ (yellow dash-dotted line) and the atomic limit for $J/U\ll 1/3$ (blue solid line) and $J/U\gg 1/3$ (red dashed line). The gray vertical line has been drawn at $J/U=1/3$. The two sketches show the fermionic configurations for MI-like (left) and HI-like (right) states.}
    \label{fig:E_0_vs_J_U_Kanamori_3_with_inset}
\end{figure}

\begin{figure}[h!]
    \centering
    \includegraphics[width=1\linewidth]{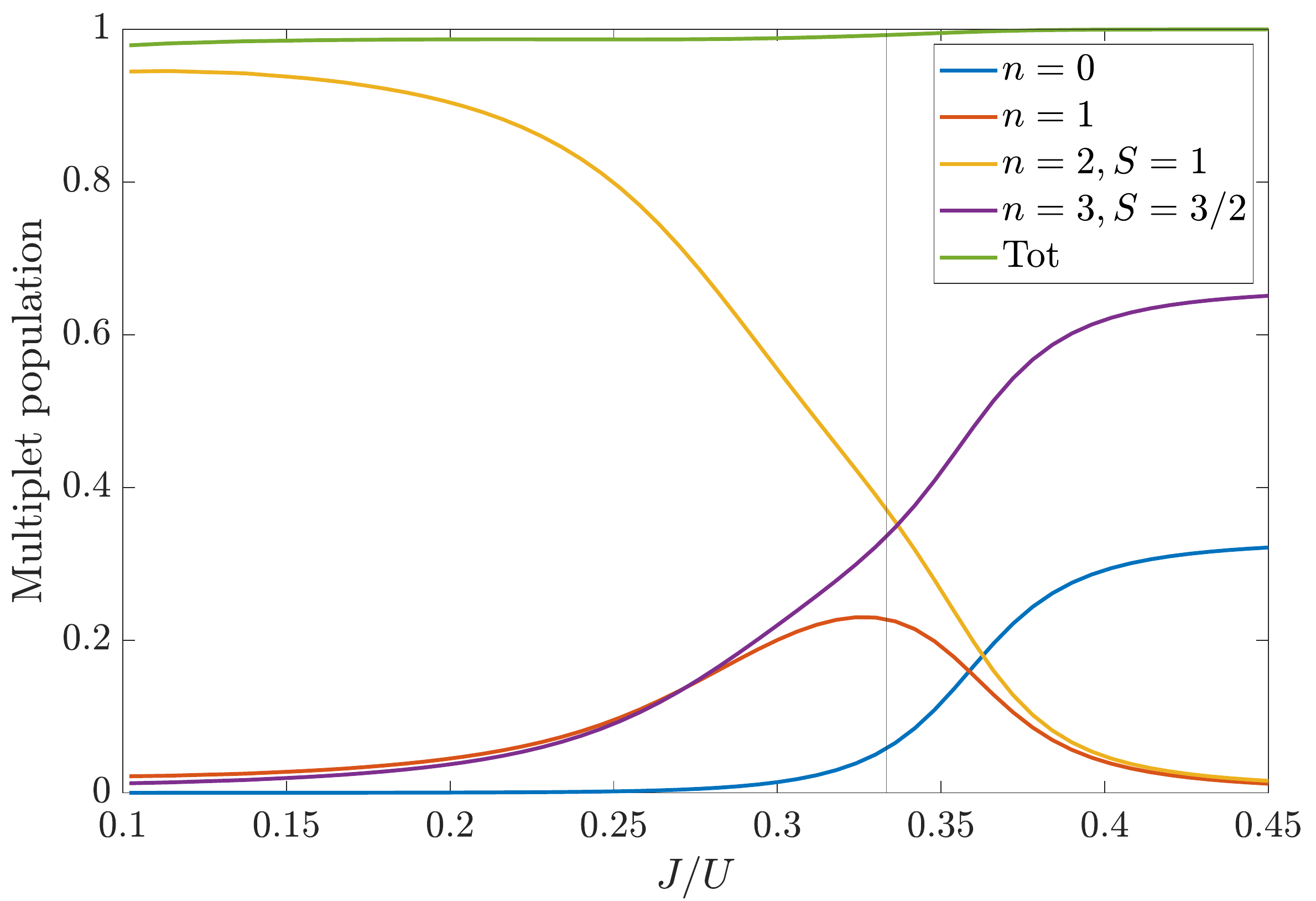}
    \caption{Multiplets' populations relevant to the ground state of Hamiltonian (\ref{eq:H_Kanamori}) as functions of the control parameter $J/U$. $t=1$, $U=22$ and 6 particles have been used. The gray vertical line has been drawn at $J/U=1/3$.}
    \label{fig:Multiplets_populations_Kanamori_3}
\end{figure}

If $J/U$ is small, the system is in a Mott insulating (MI) state, which means that there are exactly 2 fermions on each site ($n_j = 2, \, \forall\, j$). Moreover, due to the presence of a non-zero $J$, on each site we have $S_j=1$ and $L_j=1$, and so the energy reads $E_{\mathrm{MI}}=3(U-3J)$. For larger values of $J/U$ the tendency to realize high-spin configurations is so strong that it becomes energetically convenient to have different numbers of fermions on different sites. In particular, for $1/3<J/U<3/4$, the ground state features two sites with $n_j=3$, $S_j=3/2$, and $L_j=0$, while one site is empty ($n_j=S_j=L_j=0$). This state is a disproportionated Hund's insulator (HI)~\cite{Isidori_Charge_disproportionation} with energy $E_{\mathrm{HI}}=6(U-3J)$. Thus the MI and the HI are degenerate, $E_{\mathrm{MI}}=E_{\mathrm{HI}}$, for $J/U=1/3$, as shown in Fig. \ref{fig:E_0_vs_J_U_Kanamori_3_with_inset}, where we compare these limiting results with the exact ground state energy $E_0 = \langle \psi_0| H|\psi_0\rangle$ computed for a finite $U/t =22$. 
%
%
%
%
Following Ref. ~\cite{Isidori_Charge_disproportionation}, in Fig. \ref{fig:Multiplets_populations_Kanamori_3} we show the evolution, as a function of $J/U$, of the population of the most relevant atomic multiplets for $U/t = 22$. The populations are simply the sum over degenerate atomic states of $|\langle n_j, S_j |\psi_0 \rangle|^2$ where $|n_j, S_j \rangle$ is an atomic state with the corresponding quantum numbers. The results do not depend on the site because of translation invariance. We only show the most relevant configurations, which are those with a high spin, because of the finite value of the Hund's coupling.
As expected, for small values of $J/U$, the ground state is a high-spin MI and the configuration $n=2$, $S=1$ is predominant. Conversely, for large values of $J/U$, we find the two local configurations with different values of $n$ characteristic of the HI: high-spin triplets $(n=3,\,S=3/2)$ with weight $\approx 2/3$ and empty sites with weight $\approx 1/3$.
Following, in  Fig.  \ref{fig:Multiplets_populations_Kanamori_3}, the evolution of the multiplets as a function of $J/U$  at fixed and large $U$,  one can notice that, for $J/U\approx 1/3$, the multiplet $n=2,\,S=1$ falls, while multiplets $n=3,\,S=3/2$ and $n=0$ rise. In this region, an additional multiplet, $n=1$, gets populated. This is understood from the atomic limit, where, for $J/U=1/3$, the energy $E_{\mathrm{MI}} = E_{\mathrm{HI}}$ coincides with that of the configuration $n_1=1,\,S_1=1/2,\,L_1=1$; $n_2=2,\,S_2=1,\,L_2=1$; $n_3=3,\,S_3=3/2,\,L_3=0$ (or any other permutation of the site indices). In this regime, different high-spin multiplets with different local occupation are selected. We notice that the sum of the probabilities of these multiplets is persistently very close to 1, signaling that all the other configurations with lower spin are essentially irrelevant. In Sec. \ref{sub:Entropy} we will show that the region where different multiplets are populated is linked to a well-defined maximum of the associated entropy.

Finally we notice that this scenario is determined essentially by the competition between local interaction terms. In this light, we expect it to be robust to the variation of single-particle terms, including orbital-dependent hoppings and energy splittings between the atomic energies of the various orbitals, at least as long as these changes are not so large to overcome the effect of the on-site terms. For instance, lifting one orbital by an energy $\Delta$ favors an uneven occupation between orbitals. Therefore a $\Delta \gg J$ would quench the effect of $J$ introducing qualitative changes, while a reasonably smaller $\Delta$ would only introduce minor corrections.


\subsection{Conduction Properties}
\label{sub:Probing_metal}
\begin{figure}[h!]
    \centering
    \includegraphics[width=1\linewidth]{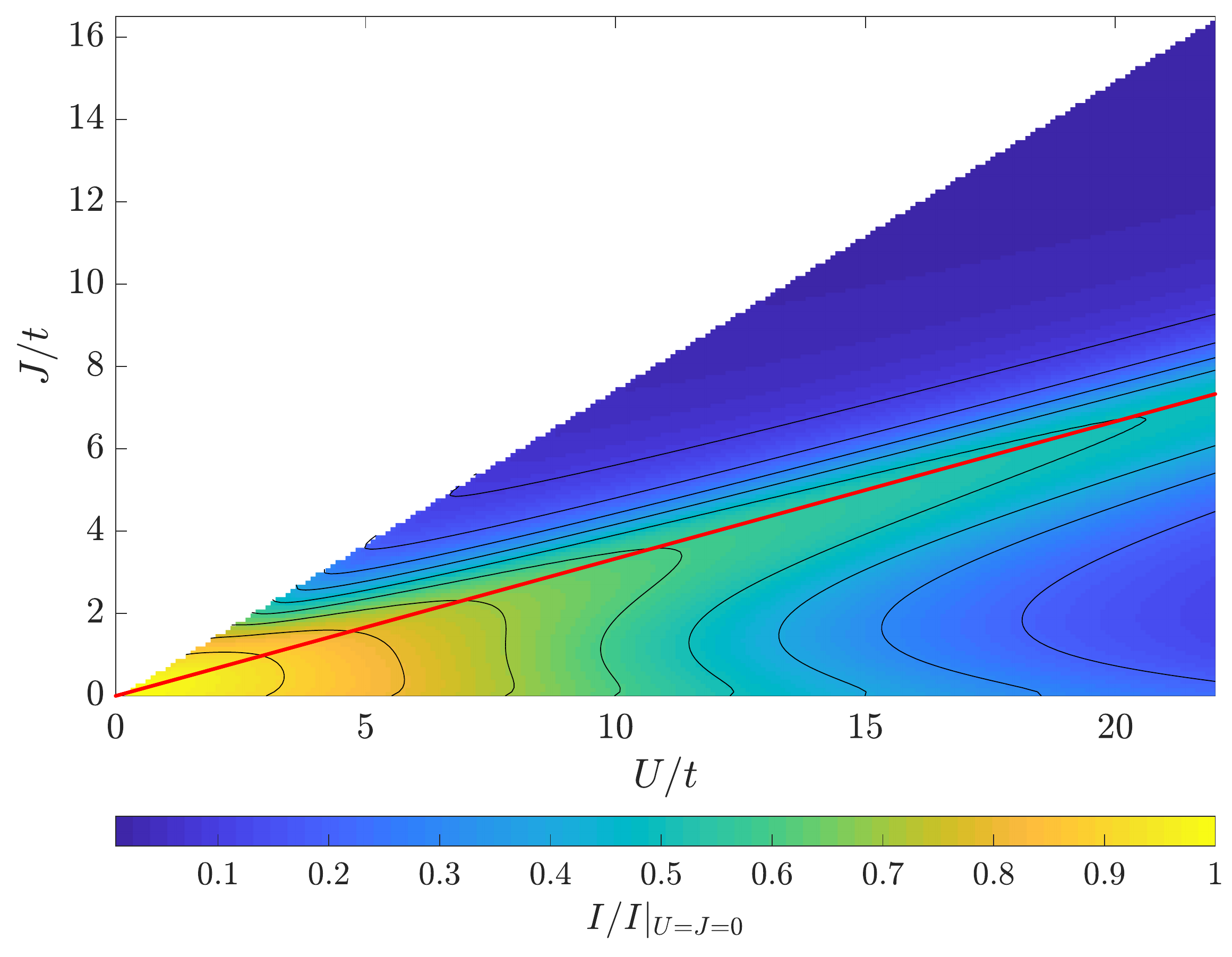}
    \caption{Expectation value of the current for $\Phi=0.1 \pi$ as a function of $U/t$ and $J/t$ for $t=1$  and 6 particles. Yellow corresponds to large values of $I$, while dark-blue corresponds to vanishing small values of $I$. The red line corresponds to $J/U=1/3$.}
    \label{fig:Piano_U_J}
\end{figure}

Even if our small system can not display real metal-insulator transitions, in this section we probe the conduction properties of the ground state to address its metallic nature in the region around $J/U = 1/3$. We follow the standard prescription~\cite{Kohn_Drude,Fye_Drude,Giamarchi_Drude,Scalapino_Drude} to compute the current by rotating the ring. In the rotating frame, the Coriolis force is formally equivalent to a magnetic flux $\Phi=2\pi m_e R^2 \Omega /\hbar $  threading the ring ($m_e$ is the fermion mass, and $R$ is the radius of the circumference where the ring-trimer is inscribed in, and $\Omega$ the angular rotation frequency) ~\cite{Arwas}. Accordingly, the hopping term of Hamiltonian (\ref{eq:H_Kanamori}) acquires a Peierls phase
\begin{equation}
    \label{eq:H_t_rotante}
    H_{\mathrm{hop}}=-t \sum_{j,a,\sigma} \left(e^{i\frac{\Phi}{L}}d_{j+1,a,\sigma}^\dagger d_{j,a,\sigma}  +\mathrm{h.c.}\right),
\end{equation}
while the interaction term (\ref{eq:H_int_Fock}) is unchanged.
Denoting with $H'$ the new Hamiltonian, and with $\left| \psi_0^\prime \right\rangle$ the corresponding ground state, we consider the expectation value of the current operator $I=\left\langle \psi_0^\prime\left| -\frac{\partial H^\prime}{\partial \Phi}\right|\psi_0^\prime \right \rangle$ in this new ground state. For small fluxes, $I$ is proportional to the Drude weight ~\cite{Giamarchi_Drude}, which is, in turn, proportional to the (singular part of the) DC electrical conductivity. Rather than a numerically unstable calculation of the conductivity in the limit of zero flux, in Fig. \ref{fig:Piano_U_J} we plot the current for a fixed small flux $\Phi=0.1 \pi$.

The results are consistent with an insulating behavior in two regions of the plane $(U/t,J/t)$. The first region is the one for large values of $U/t$ and small values of $J/t$, while the second region is found at large values of $J/t$. As discussed in Sec. \ref{sub:Rise_and_fall}, these are the MI and HI respectively.
Importantly, one can appreciate the presence of a stripe centered about the line $J/U=1/3$ where the current $I$ is persistently rather large. This is the Hund's metal region, where the simultaneous presence of different atomic multiplets (see Fig. \ref{fig:Multiplets_populations_Kanamori_3})  connected by hopping processes ensures the motion of the carriers regardless of the large values of $U$ and $J$. 


Our results follow the qualitative behaviour shown in Ref.~\cite{Isidori_Charge_disproportionation} demonstrating that the trimer captures the competition between solutions which is present in the thermodynamic limit. 

The analysis we reported so far shows that the exact results for the trimer provide a very similar picture as those obtained using RISB \cite{Isidori_Charge_disproportionation} which are in turn consistent with slave-spin \cite{De_Medici_multiorbital_correlations,Capone_Citro} and DMFT \cite{De_Medici_Janus} results.
This agreement is far from trivial since all the above approaches are defined in the thermodynamic limit, but they include only on-site dynamical correlation effects, while the solution of the trimer is numerically exact, but it is obviously limited by important finite-size effects. In this regard, the agreement is a mutual validation of the two approaches which strengthens the evidence of a correlation-resistant Hund's metal. 

\subsection{Energy spectrum and specific heat}
\label{sub:Spectrum_collapse}
\begin{figure}[h!]
    \centering
    \includegraphics[width=1\linewidth]{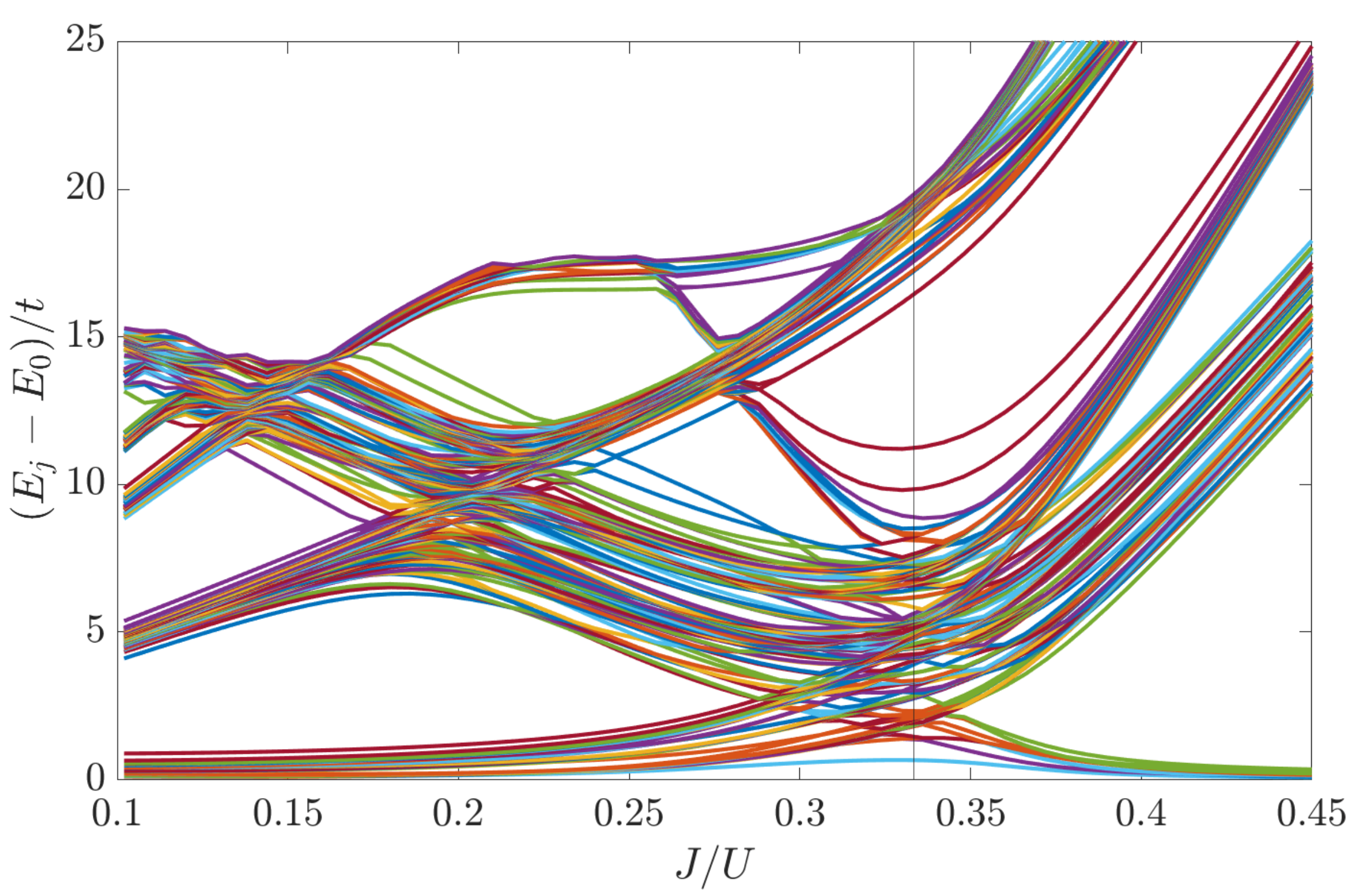}
    \caption{First 3.500 energy levels of Hamiltonian (\ref{eq:H_Kanamori}) as a function of $J/U$. $t=1$, $U=22$ and 6 particles have been used. The gray vertical line has been drawn at $J/U=1/3$.}
    \label{fig:Energy_levels_Kanamori_3}
\end{figure}

The exact numerical diagonalization of the system's Hamiltonian gives access to the full excitation spectrum. In Fig. \ref{fig:Energy_levels_Kanamori_3}, we plot the first $3.500$ energy levels $E_i=\langle \psi_i|H|\psi_i\rangle$ of Hamiltonian (\ref{eq:H_Kanamori}), as a function of the control parameter $J/U$. One can clearly notice the presence of different \textit{bundles} of energy levels, which come together or move apart upon varying $J/U$. Each bundle corresponds to a specific class of excitations. Importantly, at any given value of $J/U$, we can extract from these data valuable information about the hierarchical structure of the excitations, for example which degrees of freedom are active and which other are frozen.

 As an example, the lowest bundle which is found in the MI at $J/U=0.1$ is composed by $729=9^3$ energy levels, which is easily understood because the degeneracy of  the multiplet with $n_i=2$, $S_i=1$, and $L_i=1$ is indeed $9$ ~\cite{Isidori_Charge_disproportionation}. 
All these states are degenerate in the atomic limit and they display small splittings  for finite $t$ due to virtual hopping processes  depending on the specific arrangement of the fermions in each state. The second lowest bundle is formed by $1215=9^2\cdot 5\cdot 3$ levels. This is the number of possible states such that \textit{one} of the three sites features one minimal violation of Hund rules, namely $n_i=2,S_i=0,L_i=2$ (the degeneracy of this single-site configuration is $5$). The energy gap between this bundle and the lowest one is $2J$ ($\approx 4.4$ for $J=0.1\,U$ and for the same model parameters used in Fig. \ref{fig:Energy_levels_Kanamori_3}). With a similar reasoning, one can  verify that the  third lowest bundle, which includes  $675=9\cdot 5^2\cdot 3$ levels, corresponds to states where \textit{two} sites feature $n_i=2,S_i=0,L_i=2$ and it lies around an energy $4J$ ($\approx 8.8$ for $J=0.1\,U$ and for the same model parameters used in Fig. \ref{fig:Energy_levels_Kanamori_3}). It is only after another bundle including $243=9^2\cdot 1 \cdot 3$ levels (where $1$ is the degeneracy of single-site configuration $n_i=2$, $S_i=0$, $L_i=0$) and having energy $E_{\mathrm{MI}}+5J$, that we reach states in which the Mott condition $n_i=2$ is violated. The energy gap of these charge excitations is, for small values of $J/U$, $U-3J$. Notice that this gap depends on both $U$ and $J$, and closes exactly at $J/U=1/3$, where Hund's metallicity is found. This gap closure is clearly visible in the figure, and, indeed, it corresponds to the lowering of the energy of the charge excitation as $J/U$ grows.

In the opposite limit (right side of Fig. \ref{fig:Energy_levels_Kanamori_3}),  the lowest bundle is made up of $48=4\cdot 4 \cdot 1 \cdot 3$ levels, where $4$ is the degeneracy of the single-site configuration $n_i=3$, $S_i=3/2$, $L_i=0$, and $1$ is the degeneracy of $n_i=0$, $S_i=0$, $L_i=0$. As opposed to the case of small Hund's coupling, here the second lowest bundle already involves \textit{charge} excitations. In fact, it includes $1296=9\cdot 4\cdot 6 \cdot 6$ levels, where one factor $6$ is the degeneracy of the single-site configuration $n_1=1$, $S_1=1/2$, $L_i=1$, and the other factor $6$ represents the number of possible permutations of site indices. Notice that the gap of this bundle is of the order of $6J-2U$ ($\approx 15.4$ for $J=0.45\,U$ and for the same model parameters used in Fig. \ref{fig:Energy_levels_Kanamori_3}) and  that also this gap tends to close approaching $J/U=1/3$.

In spite of the rather complex dependence of the energy levels' structure on the control parameter $J/U$ (see Fig. \ref{fig:Energy_levels_Kanamori_3}), we have now a clear picture in which, approaching the limit $J/U = 1/3$, the charge gap collapses coming from both the MI and the HI, leading to the metallization.

\begin{figure}[h!]
    \centering
    \includegraphics[width=0.9\linewidth]{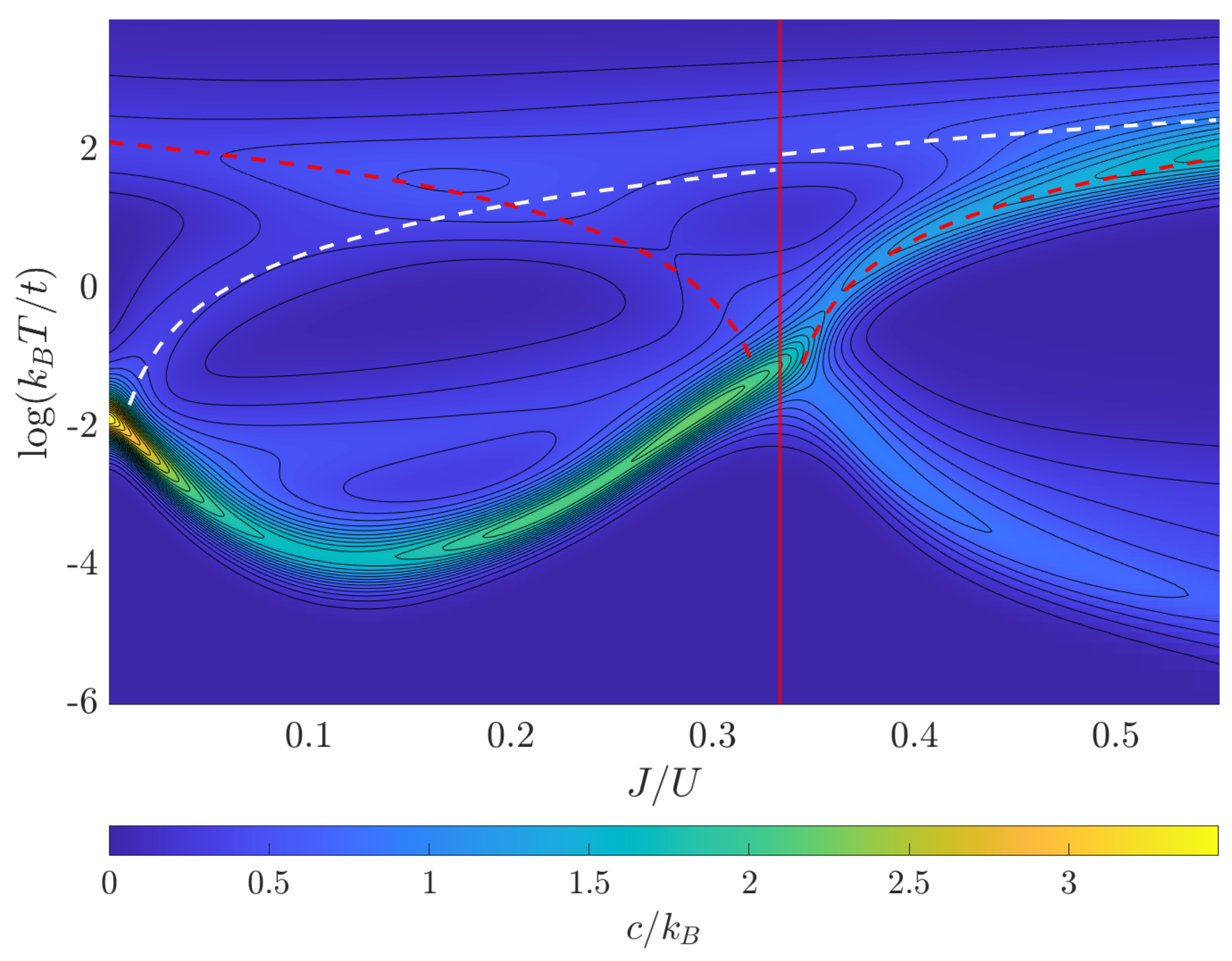}
    \caption{Specific heat as a function of $J/U$ and temperature $T$ (in logarithmic scale). Model parameters: $t=1$, $U=22$, $k_B=1$, and 6 particles. The two dashed red (white) lines correspond to simple analytical estimates of the specific heat contribution coming from charge excitations (local variations of quantum numbers $S_j$ and $L_j$) The vertical solid red line corresponds to $J/U=1/3$.}
    \label{fig:Piano_J_T_Kanamori_3}
\end{figure}

The spectrum we have discussed is naturally reflected in the thermodynamic properties. We  compute the specific heat $c=L^{-1}\partial \langle E\rangle/ \partial T$ where $\langle E \rangle = \frac{1}{Z}\sum_i E_i e^{-\frac{E_i}{k_BT}}$
is the thermal expectation value of the internal energy and $Z$ is the partition function. The result is shown in Fig. \ref{fig:Piano_J_T_Kanamori_3} in a $J/U$-$T$ plane  (where the logarithmic scale for the temperature $T$ is used for graphical clarity). For a given value of $J/U$, the specific heat (regarded as a function of $T$) features peaks where a certain class of excitations unfreezes ~\cite{Shiba_Pincus_Specific_heat,Werner_specific_heat}. Therefore, the evolution of the different ``ridges" which we observe is directly connected  with the evolution of the different bundles of energy levels in Fig. \ref{fig:Energy_levels_Kanamori_3}.

\subsection{On-site and nearest-neighbors correlation properties}
\label{sub:Correlations}

In this section, we focus on charge and spin correlation functions of the Hubbard-Kanamori trimer. This analysis extends previous investigations based on DMFT and slave-particle mean-field approaches, which only focused on \textit{local} correlators. 
We define the charge correlation functions between sites $i$ and $j$
\begin{equation}
    \label{eq:C_jj_tot}
    C_{i,j}^{\mathrm{tot}}= \langle \hat{n}_i\hat{n}_j \rangle -\langle \hat{n}_i \rangle\langle \hat{n}_j \rangle,
\end{equation}
where $\hat{n}_j = \sum_a\sum_\sigma \hat{n}_{j,a,\sigma}$ is the total density operator on site $j$. 

We can decompose the correlations in an intra-orbital and an inter-orbital contribution according to  $C_{i,j}^{\mathrm{tot}}=C_{i,j}^{\mathrm{intra}}+C_{i,j}^{\mathrm{inter}}$~\cite{Fanfarillo_Electronic_correlations}, where 
\begin{equation}
    \label{eq:C_jj_intra}
    C_{i,j}^{\mathrm{intra}}= N \left( \langle \hat{n}_{i,a}\hat{n}_{j,a} \rangle -\langle \hat{n}_{i,a} \rangle \langle \hat{n}_{j,a} \rangle\right),
\end{equation}
\begin{equation}
    \label{eq:C_jj_inter}
    C_{i,j}^{\mathrm{inter}}= N(N-1)\left(\langle \hat{n}_{i,a} \hat{n}_{j,b} \rangle -\langle \hat{n}_{i,a}\rangle\langle \hat{n}_{j,b} \rangle\right).
\end{equation}

\begin{figure}[h!]
    \centering
    \includegraphics[width=0.9\linewidth]{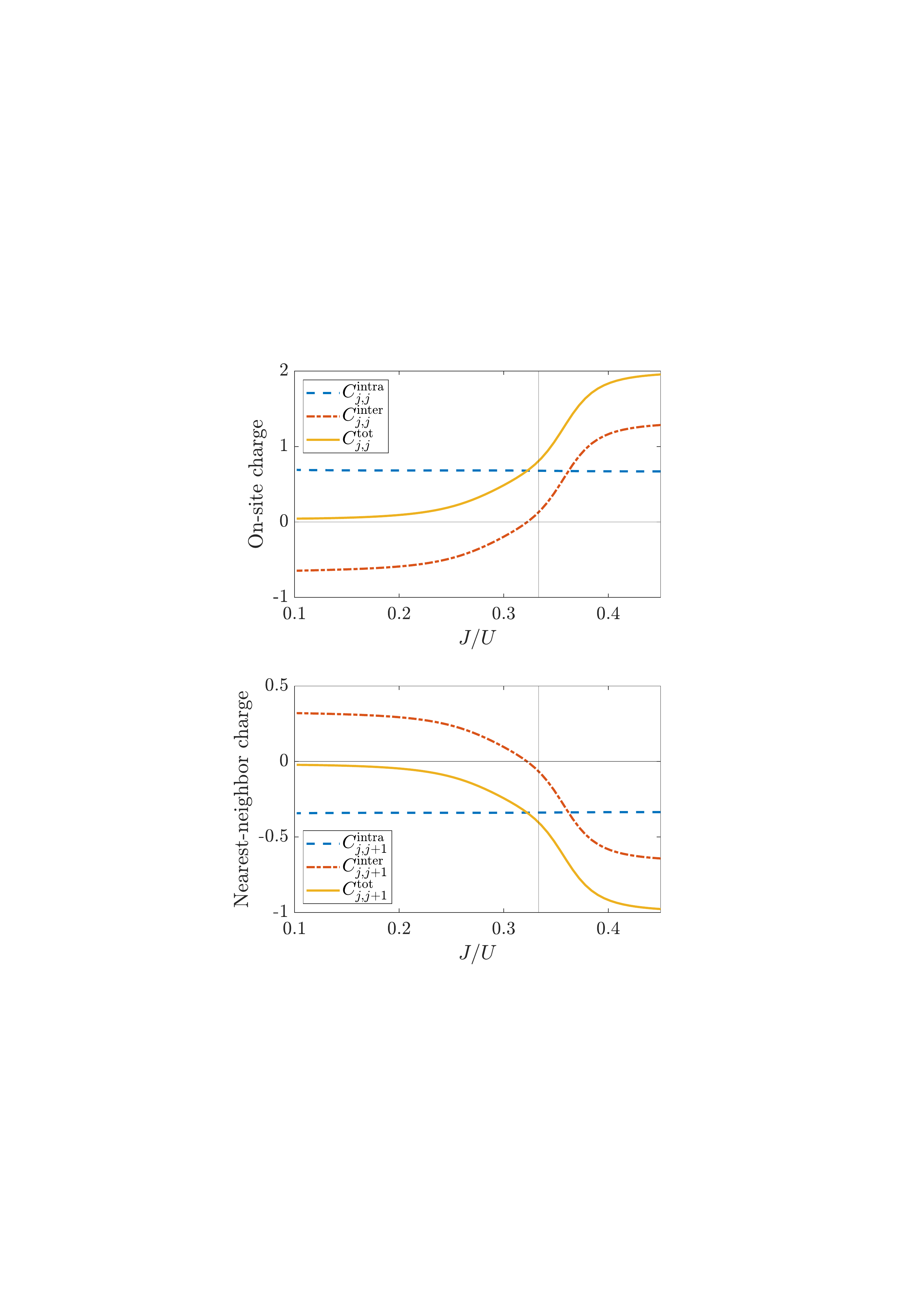}
    \caption{On-site (upper panel) and first-neighbors (lower panel) charge correlations as a function of $J/U$. $t=1$, $U=22$, and 6 particles. The vertical gray line corresponds to $J/U=1/3$.}
    \label{fig:Charge_correlations_Kanamori_3}
\end{figure}

Before discussing the results for these observables, we notice that in our small cluster we have an important constraint. The trivial relation $\langle n^2\rangle = \sum_i \langle n_i^2\rangle + 2\sum_{i<j} \langle n_i n_j\rangle$ and the fact that $\langle n^2\rangle$ is a conserved quantity implies that the sum of the on-site correlators is directly connected with the sum of the nearest-neighbor ones, which are the only non-local contributions to the sum for a three-site system. Therefore, the information about the on-site and nearest-neighbor correlation functions are not independent in our cluster. Yet, they provide us with useful physical information that complements previous studies \cite{Nonlocal_DMFT}.

Fig.~\ref{fig:Charge_correlations_Kanamori_3} illustrates the behavior of on-site (upper panel) and nearest-neighbor (lower panel) charge correlations as a function of $J$ for fixed $U=22\,t$. For both quantities, we resolve intra- and inter-orbital contributions. The behavior of the on-site charge correlation function resembles the results obtained with local mean-fields~\cite{Fanfarillo_Electronic_correlations}. The total correlations are very small in the small-$J/U$ region, where the system is a Mott insulator, and they gradually increase as $J/U$ increases and the system reaches the Hund's insulator (passing through the Hund's-metal region). We notice, in particular, that charge fluctuations are not maximal in the metallic region, but they are even larger in the Hund's insulator, a circumstance which reflects the charge disproportionation. It is also clear from the figure that the evolution as a function of $J$ of the charge correlations is entirely due to the inter-orbital component, while the intra-orbital contribution is totally unaffected by $J$ and it only depends on $U/t$. 

As expected, the results for the nearest-neighbor correlations follow a similar qualitative trend, with the total correlator vanishing in the MI and increasing (in absolute value) as $J/U$ grows. We remark that the inter-orbital correlations cross zero around the Hund's metal region for $J/U=1/3$ reflecting the decoupling between excitations in different orbitals, or orbital decoupling~\cite{De_Medici_out_lifting,De_Medici_multiorbital_correlations,Fanfarillo_Electronic_correlations,Capone_Citro}. We find therefore that the decoupling, which has been so far reported for on-site correlations, extends also to nearest-neighbor quantities, thereby strengthening its relevance.

Another important piece of information comes from the spin-spin correlation functions 
\begin{equation}
    \label{eq:M_jj_tot}
    M_{i,j}^{\mathrm{tot}}= \langle \hat{\sigma}_i\hat{\sigma}_j \rangle -\langle \hat{\sigma}_i \rangle\langle \hat{\sigma}_j \rangle
\end{equation}
where  $\hat{\sigma}_{j} = \sum_a \hat{\sigma}_{j,a}$, where $\hat{\sigma}_{j,a} = (n_{j,a,\uparrow} - n_{j,a,\downarrow})/2$.

Also in this case, the correlators can be expressed in terms of inter- and intra-orbital correlators as $ M_{i,j}^{\mathrm{tot}}=M_{i,j}^{\mathrm{intra}}+M_{i
,j}^{\mathrm{inter}}$, where 
\begin{equation}
    \label{eq:M_jj_intra}
    M_{i,j}^{\mathrm{intra}}=N\left( \langle \hat{\sigma}_{i,a}\hat{\sigma}_{j,a} \rangle -\langle \hat{\sigma}_{i,a} \rangle \langle \hat{\sigma}_{j,a} \rangle\right),
\end{equation}
\begin{equation}
    \label{eq:M_jj_inter}
    M_{i,j}^{\mathrm{inter}}= N(N-1)\left(\langle \hat{\sigma}_{i,a} \hat{\sigma}_{j,b} \rangle -\langle \hat{\sigma}_{i,a}\rangle\langle \hat{\sigma}_{j,b} \rangle\right).
\end{equation}

\begin{figure}[h!]
    \centering
    \includegraphics[width=1\linewidth]{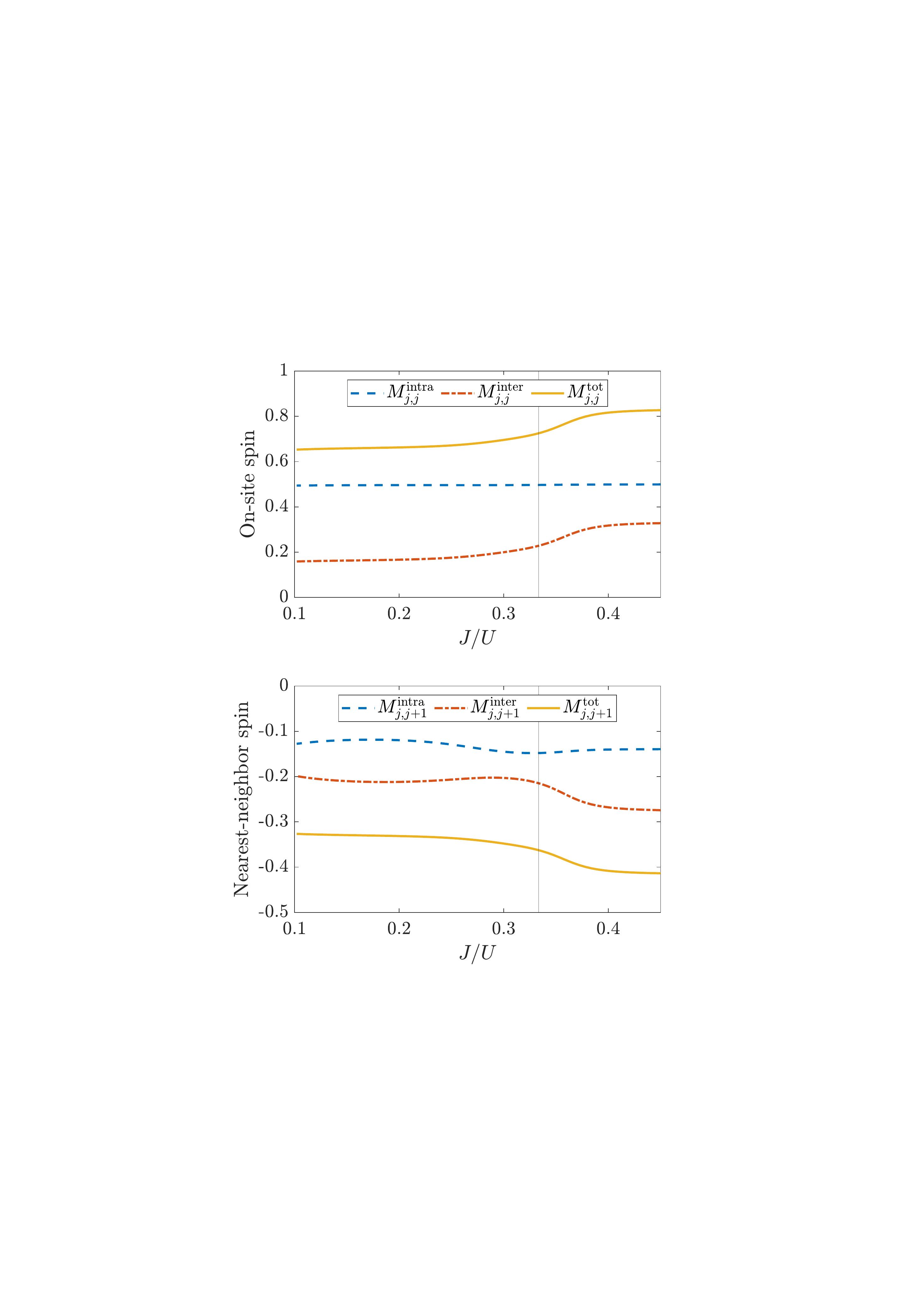}
    \caption{On-site (upper panel) and first-neighbours (lower panel) magnetic correlations with respect to the control parameter $J/U$. Model parameters $t=1$, $U=22$, and 6 particles have been used. The vertical gray line corresponds to $J/U=1/3$.}
    \label{fig:Spin_correlations_Kanamori_3}
\end{figure}

Also in this case on-site and nearest-neighbors correlators are connected by $\langle \sigma_z^2\rangle = \sum_i \langle \sigma_{i}^2\rangle + 2\sum_{i<j} \langle \sigma_i \sigma_j\rangle$.
The functional dependence of these on-site magnetic correlations on $J/U$ is illustrated in the upper panel of Fig. \ref{fig:Spin_correlations_Kanamori_3} in the regime $U/t\gg 1$. On-site spin correlations are positive and they grow with $J/U$ signalling the increased on-site magnetic moment. Since intra-orbital spin fluctuations are constant throughout the whole explored range of $J/U$, the only contribution comes from inter-orbital spin alignment due to the Hund's coupling. 

Nearest-neighbor magnetic correlations are shown in the lower panel of Fig. \ref{fig:Spin_correlations_Kanamori_3}. The total correlator $M_{j,j+1}^{\mathrm{tot}}$ is negative, signalling antiferromagnetic spin correlations between the large local magnetic moments. The absolute value grows with $J/U$ without any anomaly when the Hund's metal region is reached and crossed. An antiferromagnetic ordering is found also in the Hund's insulator. In this case the intra-orbital correlations are negative and they have a mild dependence on $J/U$ which combines with the larger dependence of the inter-orbital terms to provide the final result.


\section{The SU(3) Hubbard trimer with patterned potential}
\label{sec:Hubbard_trimer}

In this section, we focus on a fermionic SU(3) Hubbard model with a suitable patterned potential which favours a charge-ordered state. The Hamiltonian reads 
$$
  H=-t\sum_{j=1}^L \sum_{a=1}^N\left(d^\dagger_{j,a}d_{j+1,a}+\mathrm{h.c.}\right)
$$
\begin{equation}
    \label{eq:H_SU_3}
    + \frac{U}{2}\sum_{j=1}^L\hat{n}_j(\hat{n}_j-1) + \sum_{j=1}^L \mu_j \hat{n}_j,
\end{equation}
where operator $d^\dagger_{j,a}$ creates a fermion with flavor $a=1,2,3$ on site $j$, and operator $\hat{n}_j:=\sum_{a=1}^N \hat{n}_{j,a}$ counts the number of fermions at site $j$. Here, $N=3$ is the number of flavors, and $L=3$ is the number of sites, $t$ represents the hopping, $U$ is the Hubbard interaction and $\mu_j$ a site-dependent potential corresponding to the presence of a superlattice. We will assume two sites with the same energy and one at a higher energy $\mu_1=-\mu$, $\mu_2=0$, $\mu_3=-\mu$ ($\mu >0$). The physics of this model is ruled by the competition between $U$, which penalizes local occupancies different from the global average density, and $\mu$, which favours the occupation of the low-energy sites over the high-energy one. A similar pattern can also be realized in larger lattices formed by periodic repetitions of the three-site cluster. In this sense, also in this case, the trimer can be seen as the building block of a quantum cluster theory for a larger system.

The model can be seen as a three-flavour/three-site version of the ionic Hubbard model~\cite{Fabrizio_ionic,Wilkens_ionic,Japaridze_ionic,Manmana_ionic,Batista_ionic,Torio_ionic,Chattopadhyay_ionic,Garg_ionic,Paris_ionic}, where the local Hubbard repulsion competes with a staggered potential. This leads to a competition between Mott and charge-density-wave insulators, separated in the phase diagram by a narrow stripe where, depending on the dimensionality, a  metal~\cite{Bouadim_metallic_ionic_2D} or a bond-order-wave phase have been reported, the latter being characterized by a staggered kinetic-energy on the bonds~\cite{Sengupta_BOW}.
The ionic Hubbard model has been recently realized in a Fermi gas loaded into a honeycomb optical lattice with a staggered energy offset~\cite{Messer_ionic_ultracold_atoms}. The two insulating phases were observed, but the bond-order-wave phase still eludes experimental detection~\cite{Loida_Bond_order_wave}. 

This kind of system, where the three-site unit is periodically repeated in space, can be experimentally realized by means of a multicomponent fermionic quantum gas, for example $^{6}\mathrm{Li}$~\cite{Ottenstein_Li_6,Huckans_Li_6}, $^{87}\mathrm{Sr}$~\cite{Strontium87}, or  $^{173}\mathrm{Yb}$~\cite{Ytterbium173} preparing a balanced mixture of three species with different nuclear spin. The second step is to introduce an optical lattice which can be partitioned in three sublattices such as the triangular~\cite{Triangular_lattice} or the Kagome lattice~\cite{Kagome_lattice} in two dimensions, or a simple one-dimensional chain \cite{mancini2015observation}, thus realizing a SU($N$) Hubbard model~\cite{Gorshkov}. If the laser is used to build one of these lattices, we would realize a model with uniform single-particle potential. In order to introduce the patterned potential where one site out of three has a higher energy, one can superimpose a second optical lattice with a larger wavelength which selects only one of the three sublattices~\cite{Messer_ionic_ultracold_atoms, Superlattice}.

\subsection{From a Mott-like insulator to a band-like insulator through metal-like states}
\label{sub:Fro_MI_to_BI}

\begin{figure}[h!]
    \centering
    \includegraphics[width=1\linewidth]{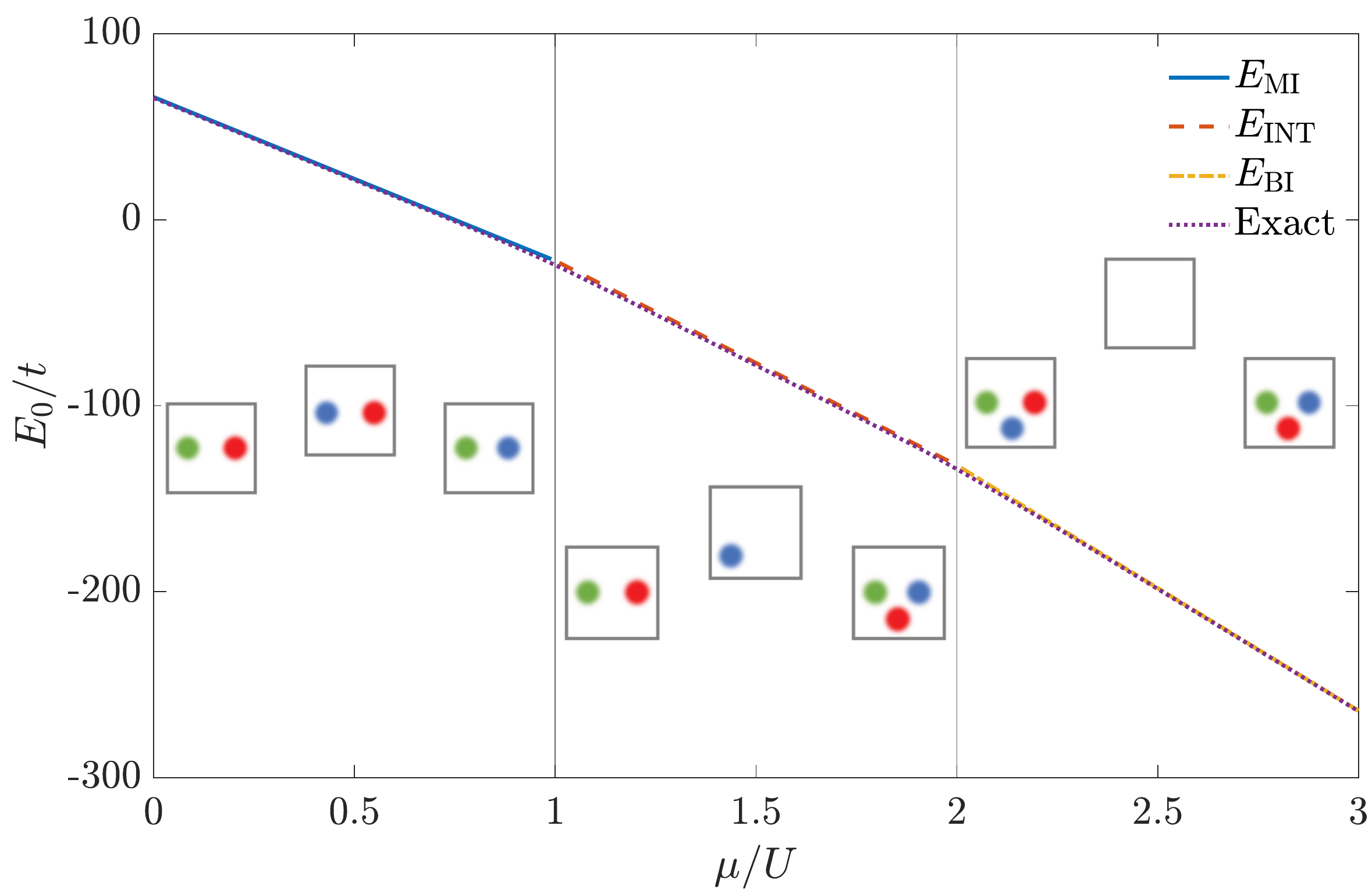}
    \caption{Ground state energy of a three-flavor Hubbard trimer with staggered potential hosting 6 fermions for $t=1$, $U=22$ (purple dotted line). Atomic estimates for $\mu/U<1$ (blue solid line), $1<\mu/U<2$ (red dashed line), and $\mu/U>2$ (yellow dash-dotted line) are also shown. The three atomic configurations are sketched in the corresponding regions. The gray vertical lines have been drawn at $\mu/U=1$ and $\mu/U=2$.}
    \label{fig:E_0_vs_mu_U_SU_3_with_inset}
\end{figure}

We start our study from the atomic limit ($t=0$) of the Hamiltonian (\ref{eq:H_SU_3}) in the presence of $2L=6$ fermions. In contrast with the Hubbard-Kanamori model, we find three different regimes (see Fig. \ref{fig:E_0_vs_mu_U_SU_3_with_inset}). For $0<\mu/U<1$, the ground state describes a Mott insulator with 
exactly two fermions per site (see the left part of the upper panel of Fig. \ref{fig:Multiplets_populations_SU_3}), which cannot hop due to the large Coulomb repulsion $U$. The energy of this configuration is $E_{\mathrm{MI}}=3U-4\mu$. On the other hand, for $\mu/U>2$, the non-uniform potential prevails over the Coulomb interaction and it is energetically convenient to pack all the fermions in the two low-energy sites, thus manifestly violating Mott's condition (see the right part of the upper panel of Fig. \ref{fig:Multiplets_populations_SU_3}). This configuration, whose energy is $E_{\mathrm{BI}}=6U-6\mu$ can be regarded as the atomic version of a band insulator (BI) (in analogy to the band-insulating phase of the well-known ionic Hubbard model ~\cite{Bouadim_metallic_ionic_2D}). Indeed, if we consider a large lattice with a finite hopping, the three atomic levels will broaden into bands. For small hopping amplitudes, the bands arising from the low-energy sites will be fully occupied while the band originating from the high-energy site will be empty. These two solutions are conceptually connected with the Mott and Hund's insulators found in the three-orbital Hubbard-Kanamori model.

While the two solutions of the previous model become degenerate on a line, here we find a whole intermediate solution which is stable in the range $1<\mu/U<2$. Here, the competition between $U$ and $\mu$ results in a set of degenerate ground states such that the high-energy site hosts one fermion, while the low-energy sites host the remaining 5 fermions (see the central part of the upper panel of Fig. \ref{fig:Multiplets_populations_SU_3}). 
The total energy of this intermediate configuration, in the atomic limit, reads $E_{\mathrm{INT}}=4U-5\mu$ and, as illustrated in Fig. \ref{fig:E_0_vs_mu_U_SU_3_with_inset}, one can verify that $E_{\mathrm{MI}}=E_{\mathrm{INT}}$ for $\mu/U=1$, and that $E_{\mathrm{INT}}=E_{\mathrm{BI}}$ for $\mu/U=2$. 

In the following, we discuss the fate of these solutions once a finite hopping is included and their connection with the Hund's metal.

\begin{figure}[h!]
    \centering
    \includegraphics[width=1\linewidth]{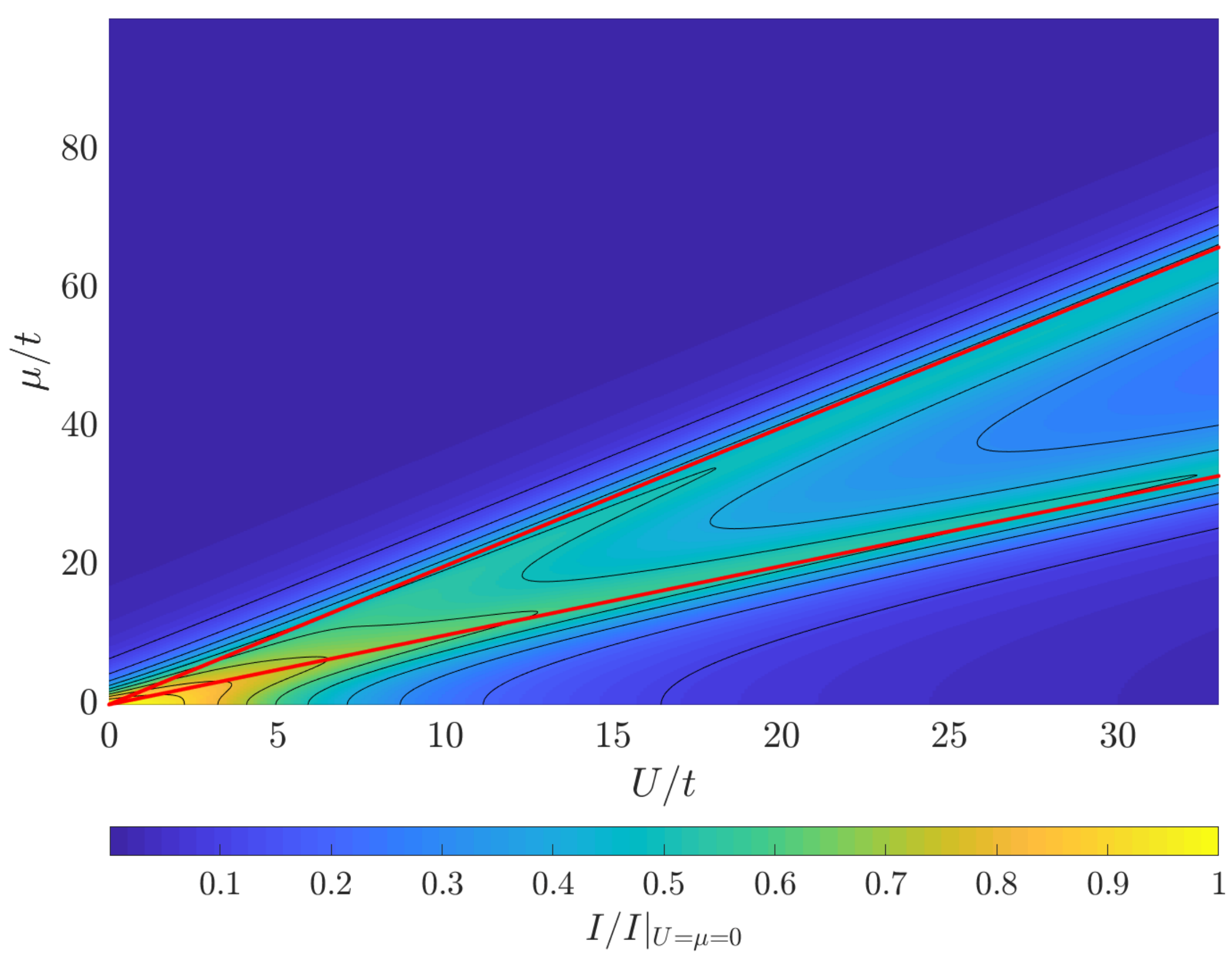}
    \caption{Expectation value of the current operator, as a function of the control parameters $U/t$ and $\mu/t$ for 6 fermions and $\Phi=0.4\, \pi$. Yellow corresponds to large values of $I$, while dark-blue corresponds to a vanishing small value of $I$. The red lines correspond to $\mu/U=1$ and $\mu/U=2$.}
    \label{fig:Piano_U_mu}
\end{figure}

\subsection{Conduction properties}

We begin the characterization of the model for finite value of the hopping by computing the current as described in  Sec. \ref{sub:Probing_metal}. 
The results are reported in Fig. \ref{fig:Piano_U_mu}. For $U/t\gg 1$ and small values of $\mu/t$, we recover the familiar Mott insulator, in which the current is suppressed in order to avoid creating triple occupancies. On the other hand, for $\mu/U \gg 2$, we have a band insulator with fully occupied low-energy sites and empty high-energy site, a configuration which inhibits the current as well. 

Interestingly, and in contrast with the Hubbard-Kanamori system, the intermediate region between the two limiting lines $\mu=U$ and $\mu=2U$ does not host a metallic region. When we approach and cross the two lines $\mu=U$ and $\mu=2U$, we have an enhancement of the current, while the intermediate region between the two lines appears to host a metal only for small values of $U$ and $\mu$. These results suggest that an interaction-resistant metal is only realized close to the boundary lines found in the atomic limit, while the intermediate region appears to undergo a Mott transition as $U/t$ increases at fixed $\mu/U$. However, the stabilization of a metal for very large values of $U$ and $\mu$ along the two boundary lines emerges as the counterpart of the Hund's metal for the present SU(3) model. In the next section, we explore in more depth the connection between the two results.

\subsection{Local configurations in the SU(3) Hubbard model with patterned potential}
\label{sub:Shallower_vs_deeper}

\begin{figure}[h!]
    \centering
    \includegraphics[width=1\linewidth]{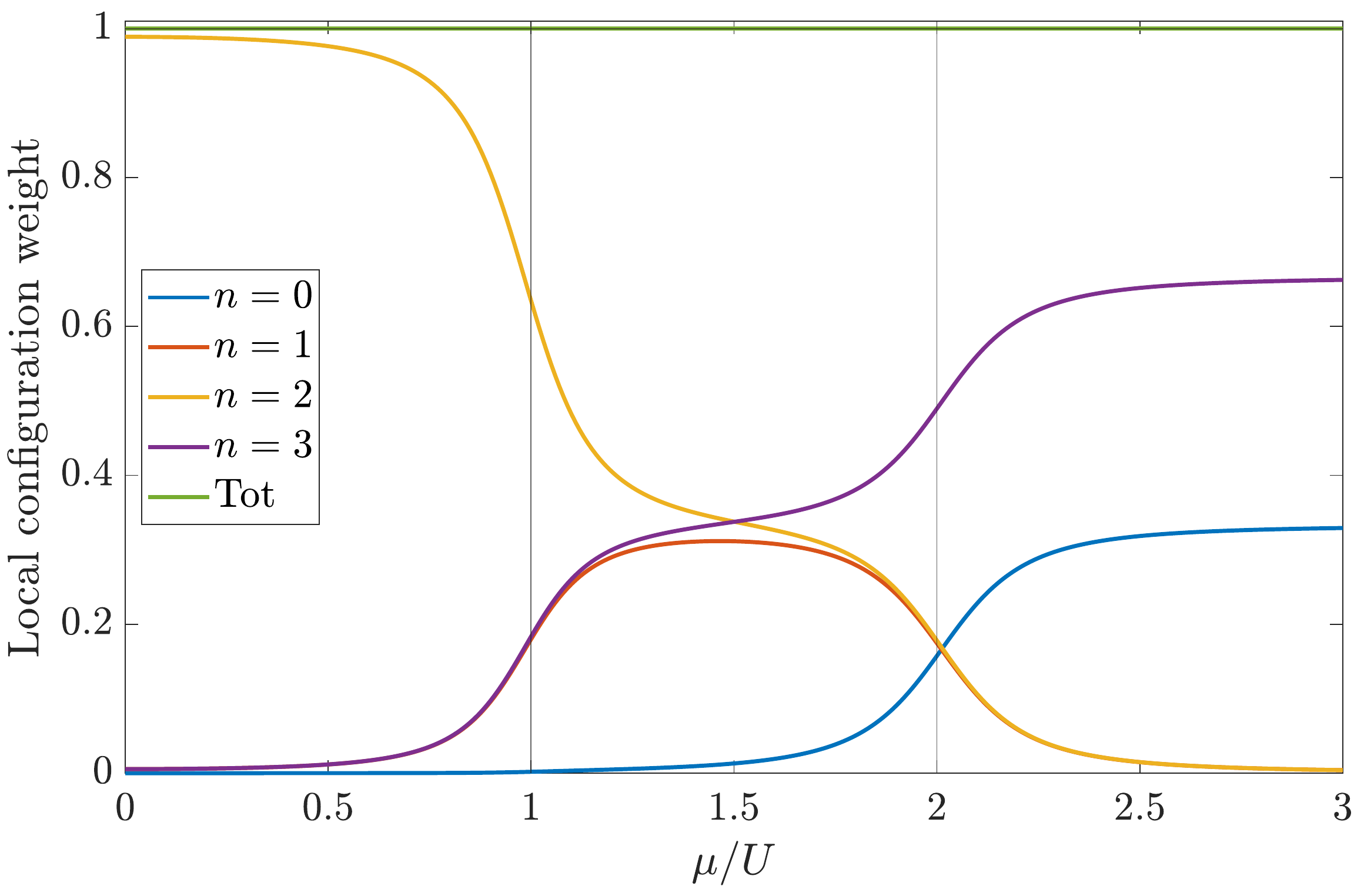}
     \includegraphics[width=1\linewidth]{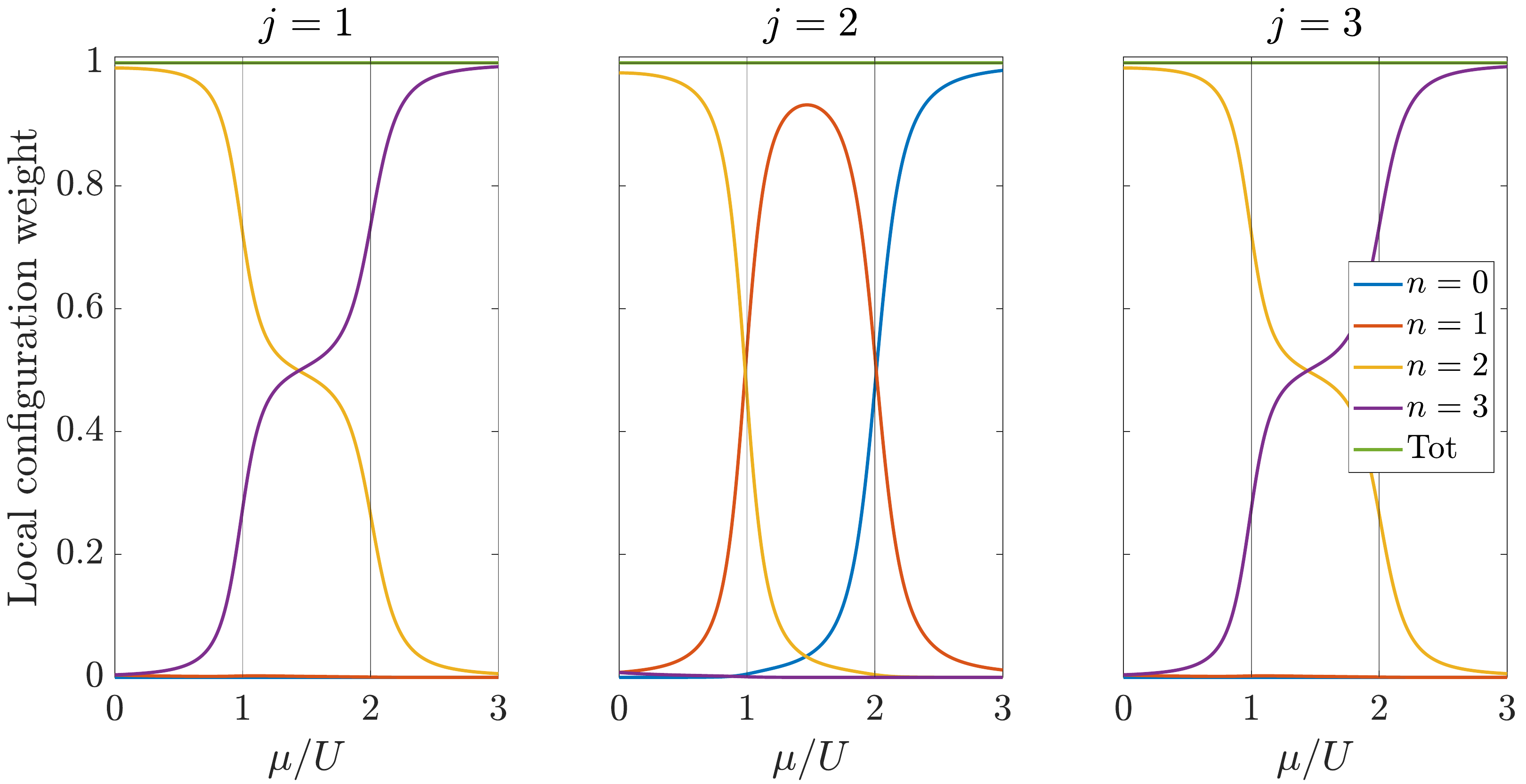}
   \caption{The top panel shows the occupation of the atomic states in the ground state of the Hamiltonian (\ref{eq:H_SU_3}) as a function of  $\mu/U$. $t=1$, $U=22$ and 6 particles have been used. The gray vertical lines have been drawn at $\mu/U=1$ and $\mu/U=2$.
      The bottom panels show the site-resolved occupations for the three sites.}
    \label{fig:Multiplets_populations_SU_3}
\end{figure}

In the upper panel of Fig.~\ref{fig:Multiplets_populations_SU_3}, we report the population of states with fixed number of fermions per site in the SU(3) Hubbard model with patterned potential in the strong-coupling regime. Given the nature of the atomic states of this model, we do not need to disentangle the contribution between different multiplets with the same local occupation. Our results clearly highlight that the solutions for finite $t$ are connected with the atomic solution. In particular, it is evident that we obtain a Mott insulator (where all the sites have $n=2$) for small values of $\mu/U$, and a band insulator [with two filled sites ($n=3$) and one empty site ($n=0$)] for large $\mu/U$. The intermediate region is instead characterized by a nearly uniform probability distribution of local configurations with $n=1$, $n=2$ and $n=3$, while the one with $n=0$ is suppressed. Interestingly, comparing the present results with the plot of the current of Fig. \ref{fig:Piano_U_mu}, the maximum value of the current is not obtained in the region where the local configurations are similar, but rather close to the boundaries of the intermediate region. This result may appear surprising since a metal is a state characterized by large density fluctuations. 

We can understand this discrepancy noting that the present SU(3) Hubbard model is not translationally invariant because of the different local potential in the three sites. A more insightful picture of the local configurations and their relation with the metallic behavior can be obtained by showing the local occupation at each site, as we do in the lower panels of Fig. \ref{fig:Multiplets_populations_SU_3}. It is clear that, for large $\mu/U$, the low-energy sites 1 and 3 are completely filled, while the high-energy site 2 is practically empty (see Sec. \ref{sub:Fro_MI_to_BI}), while for small $\mu/U$ all the three sites host 2 fermions.

On the other hand, in the middle of the intermediate region (consider, for example, $\mu \simeq 1.5\, U$), the high-energy site hosts $1$ fermion, while the low-energy sites are occupied by the remaining five fermions (each site with an equal probability to have two or three fermions). This configuration is not favourable for conduction because the hopping processes involving the high-energy site have a large energetic cost, so that the only allowed hopping processes are those connecting the low-energy sites. 

When we approach the boundaries of the intermediate region, for example for $\mu \simeq U$, the low-energy sites have a finite amplitude to host two or three fermions, while the high-energy site can host one or two fermions. This pattern of local configurations allows instead for hopping processes as long as the hopping is finite, giving rise to a situation which closely resembles the Hund's metal.

This analysis clarifies the physical origin of the behavior of the current. In the intermediate region $1<\mu/U<2$,  we certainly have hopping processes connecting the two low-energy sites, while the high-energy one remains disconnected as long as the hopping is not large enough to overcome the energy difference. According to the geometry and the dimensionality of the lattice, this may lead to metallic or insulating phases. In our small cluster, we clearly find a reduction of the conductivity as $U$ increases along all the lines with constant $\mu/U$ within the window $1<\mu/U<2$. The residual current is merely associated with the hopping between the low-energy sites. 

Along the two lines in the $(U/t,\mu/t)$ plane corresponding to $\mu/U=1$ and to $\mu/U=2$, where the intermediate phase becomes degenerate with one of the two limiting solutions in the atomic limit, we find, instead, a metallic solution which exploits the competition between two different tendencies controlled by $U$ and $\mu$. 


\subsection{Entropy and multiplet population}
\label{sub:Entropy}
We conclude our investigation of the analogies and differences between the two models by computing the entropy, which is in turn directly connected with the distribution of the local configurations reported in  Fig. \ref{fig:Multiplets_populations_Kanamori_3} and Fig. \ref{fig:Multiplets_populations_SU_3}.
As anticipated in Sec. \ref{sub:Rise_and_fall}, it is possible to compute the entropy as
\begin{equation}
    \label{eq:mathcal_S}
    \mathcal{S}= -\sum_\ell p_\ell \log(p_\ell),
\end{equation}
where $p_\ell$ is the population of the $\ell$-th atomic state. 

\begin{figure}
    \centering
    \includegraphics[width=0.8\linewidth]{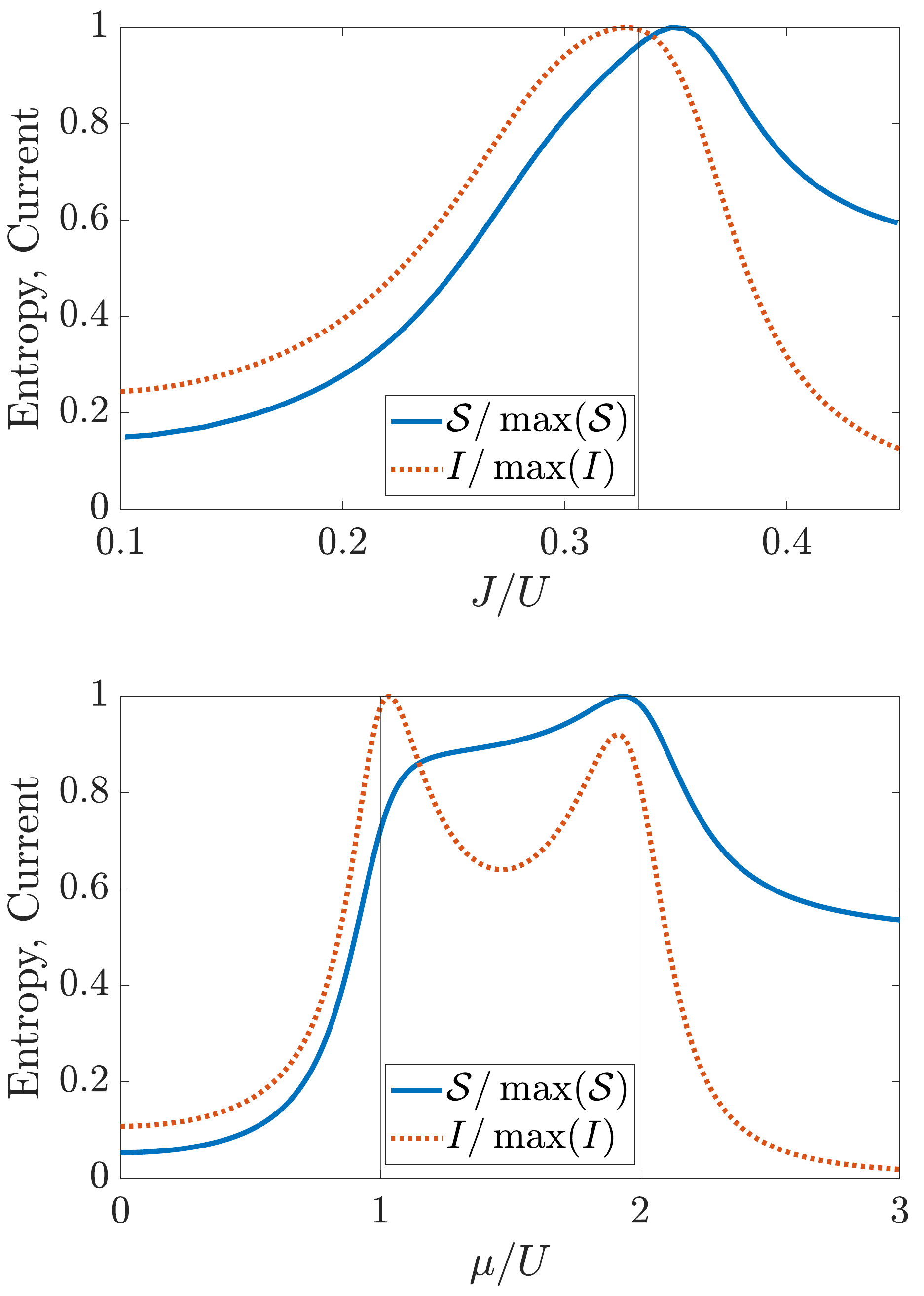}
    \caption{Comparison between the entropy associated with the occupation of atomic states $\mathcal{S}$ and the current $I$ for the Hubbard-Kanamori model (upper panel) and the SU(3) Hubbard model (lower panel). For both panels, model parameters $t=1$, $U=22$ have been used, and the data have been normalized to $1$ in order to allow for a direct comparison. }   
    \label{fig:Entropy_of_multiplets}
\end{figure}

In Fig. \ref{fig:Entropy_of_multiplets} we show, for both models, the evolution of the entropy and of the current along the usual cuts for $U/t =22$. In the case of the Hubbard-Kanamori trimer (upper panel), $\mathcal{S}$ and $I$ increase in a very similar way in the region $J/U < 1/3$ and they reach relatively close maxima in the Hund's metal region before dropping when the Hund's insulator regime is reached. In this case the best metallic behavior is found  where various atomic states are democratically populated, so that the entropy is maximized. For $J/U\gtrsim 1/3$, $\mathcal{S}$ converges towards $0.64 \approx -2/3\log(2/3)-1/3\log(1/3)$, while $I$ drops towards $0$. This happens because the two multiplets which remain populated and determine a non-zero $\mathcal{S}$ are not compatible with hopping processes (and hence cannot support conductance). 

For the SU(3) Hubbard trimer (lower panel of Fig. \ref{fig:Entropy_of_multiplets}), we find a similar behavior in the two external regions. Within the Mott insulator the entropy and the current increase at the same rate. In the intermediate region we find a different behavior, consistent with the previous observations. The current is maximized close to $\mu=U$ and $\mu=2U$ while the entropy remains large in the whole intermediate window. This is due to the fact that the atomic states selected in the middle of the intermediate region are not compatible with the conduction. The large $\mu/U$ behavior is similar to the large $J/U$ region of the Hubbard-Kanamori model with a finite entropy associated with a residual degeneracy which does not lead to a finite current because the degenerate states are not connected by hopping.

\section{Concluding remarks}
\label{sec:Concluding_Remarks}
In this manuscript, we have studied the general mechanism that stabilizes an interaction-resistant metal, i.e. a metal which survives to large values of repulsion at integer fillings. In particular, we have uncovered the existence of a similar metallic phase in two rather different models of strongly correlated fermions.

Namely, we studied and compared a three-orbital Hubbard model and a three-component SU(3) Hubbard model with a three-site patterned potential. 
In both models, the standard Hubbard repulsion, which tends to stabilize a Mott insulator, competes with a term of the Hamiltonian which favours a different state with inhomogeneous density distribution, namely the Hund's coupling in the first model and a non-uniform single-particle potential in the second. In both cases, the competition between the two terms of the Hamiltonian leads to different insulating solutions which are separated in the respective phase diagrams by families of states which exhibit persistent metallicity,  even in the presence of strong Coulomb repulsions, hence the name ``interaction-resilient metals". In the case of the Hubbard-Kanamori model, this metallic state has been recently labelled Hund's metal.

The exact results on the small cluster allow us to provide information on the nearest-neighbor correlation functions of the model, which have not been discussed using the above-mentioned approaches. In particular, we have demonstrated that, in the Hund's metal region, the nearest-neighbor inter-orbital charge correlations vanish just like the onsite components, strengthening the relation between the Hund's metal and the effective decoupling between orbitals. Moreover we have found that the nearest-neighbor spin correlators are always negative implying a tendency towards antiferromagnetic ordering in all the regions of the phase diagram, including the Hund's metal and the Hund's insulator.

We have supplemented previous studies with a detailed analysis of the fingerprint associated to the formation of the Hund's metal in the evolution of the many-body energy spectrum and in the temperature dependence of the specific heat. 

In the second part of the manuscript, we have performed a similar analysis for the SU(3) Hubbard model with a three-site energy pattern (one site out of three with a higher energy).
Here, we find a slightly richer phase diagram, which is marked by two different lines along which metallic solutions outlive large interactions, in analogy with the Hund's metal. This result is understood by inspecting the probability distribution of different local configurations in the ground state, and is connected with the similar physical scenario emerging within the Hubbard-Kanamori model. An investigation of the entropy and its comparison with the conductivity strengthens the connection between the two models and the identification of the mechanism behind the stabilization of the discussed interaction-resilient metals.

We have therefore provided a strong evidence that an interaction-resistant metallic state is not peculiar of the Hund's physics, where it has been widely discussed, but it is a more general feature which is present for a wide class of models featuring competing insulating states associated with different local configurations. Tuning the parameters to make the insulating states degenerate or nearly degenerate, a correlation-resistant metal exists as long as the local configurations corresponding to the two insulators are connected by hopping processes. 

In this work, we considered a minimal three-site cluster to 
explore the differences and similarities between the two models. For the Hubbard-Kanamori model, our trimer reproduces the scenario obtained with other approaches including DMFT, RISB, and slave-spin mean-field. This agreement between different approaches represents a mutual validation of the different methods and confirms that the existence of the Hund's metal is an intrinsic feature of the multiorbital Hubbard model.

Moreover, exact result for small clusters can be used to reconstruct the properties of infinite lattices using quantum cluster methods such as cluster perturbation theory \cite{Cluster-Perturbation-Theory}, variational cluster approximation \cite{Potthoff}, or cluster extensions of DMFT \cite{CDMFT,DCA}. The discussed results represent, in this perspective, the basic building block from which the lattice physics can be built. These approximations are particularly accurate for the strongly correlated insulating solutions, which have an essentially local character, but they should not alter significantly the shape of the phase diagram including the region where the interaction-resistant metal is stable.



\section*{Acknowledgements}
\label{sec:Acknowledgements}
The authors would like to thank L. de' Medici, L.F. Tocchio, A. Isidori and L. Livi for fruitful discussions. We acknowledge financial support from MIUR through the PRIN 2017 (Prot. 20172H2SC4 005) programs and Horizon 2020 through the ERC project FIRSTORM (Grant Agreement 692670).

\end{document}